\documentclass[pdflatex,sn-mathphys-num]{sn-jnl}

\usepackage{graphicx}%
\usepackage{multirow}%
\usepackage{amsmath,amssymb,amsfonts}%
\usepackage{amsthm}%
\usepackage{mathrsfs}%
\usepackage[title]{appendix}%
\usepackage{xcolor}%
\usepackage{textcomp}%
\usepackage{manyfoot}%
\usepackage{booktabs}%
\usepackage{algorithm}%
\usepackage{algorithmicx}%
\usepackage{algpseudocode}%
\usepackage{listings}%
\usepackage{soul}

\usepackage{gensymb}
\usepackage{multirow}

\usepackage{booktabs}

\usepackage{caption}
\usepackage{array}
\usepackage{svg}

\theoremstyle{thmstyleone}%
\theoremstyle{thmstyletwo}%

\theoremstyle{thmstylethree}%

\newcounter{extendedfigure} 
\newenvironment{extended_figure}[1][]
{
    \begin{figure}[h!]
    \centering
    \renewcommand{\figurename}{Extended Data Fig.}
    \stepcounter{extendedfigure} 
    \renewcommand{\thefigure}{\arabic{extendedfigure}} 
}
{
    \end{figure}
    \renewcommand{\figurename}{Figure}
    \renewcommand{\thefigure}{\arabic{figure}}
}

\newcounter{extendedtable} 
\newenvironment{extended_table}[1][]
{
    \begin{table}[h!]
    \centering
    \captionsetup{font=small, labelfont=bf, labelsep=period}
    \renewcommand{\tablename}{Extended Data Table}
    \stepcounter{extendedtable} 
    \renewcommand{\thetable}{\arabic{extendedtable}} 
}
{
    \end{table}
    \renewcommand{\tablename}{Table}
    \renewcommand{\thetable}{\arabic{table}}
}

\begin{document}

\title[Article Title]{
FetalCLIP: A Visual-Language Foundation Model for Fetal Ultrasound Image Analysis
}


\author[1]{\fnm{Fadillah} \sur{Maani}}
\equalcont{Contributed Equally.}

\author[1]{\fnm{Numan} \sur{Saeed}}
\equalcont{Contributed Equally.}

\author[1]{\fnm{Tausifa Jan} \sur{Saleem}}

\author[1]{\fnm{Zaid} \sur{Farooq}}

\author[2]{\fnm{Hussain} \sur{Alasmawi}}

\author[3]{\fnm{Werner} \sur{Diehl}}

\author[3]{\fnm{Ameera} \sur{Mohammad}}

\author[3]{\fnm{Gareth} \sur{Waring}}

\author[3]{\fnm{Saudabi} \sur{Valappil}}

\author[3]{\fnm{Leanne} \sur{Bricker}}

\author*[1]{\fnm{Mohammad} \sur{Yaqub}}
\email{mohammad.yaqub@mbzuai.ac.ae}

\affil[1]{\orgdiv{Department of Computer Vision}, \orgname{Mohamed bin Zayed University of Artificial Intelligence}, \orgaddress{\city{Abu Dhabi}, \country{UAE}}}

\affil[2]{\orgdiv{Department of Machine Learning}, \orgname{Mohamed bin Zayed University of Artificial Intelligence}, \orgaddress{\city{Abu Dhabi}, \country{UAE}}}

\affil[3]{\orgname{Corniche Hospital, Abu Dhabi Health Services Company (SEHA)}, \orgaddress{\city{Abu Dhabi}, \country{UAE}}}  


\abstract{Foundation models are becoming increasingly effective in the medical domain, offering pre-trained models on large datasets that can be readily adapted for downstream tasks. Despite progress, fetal ultrasound images remain a challenging domain for foundation models due to their inherent complexity, often requiring substantial additional training and facing limitations due to the scarcity of paired multimodal data. To overcome these challenges, here we introduce \textbf{FetalCLIP}, a vision-language foundation model capable of generating universal representation of fetal ultrasound images. \textbf{FetalCLIP} was pre-trained using a multimodal learning approach on a diverse dataset of 210,035 fetal ultrasound images paired with text. This represents the largest paired dataset of its kind used for foundation model development to date. This unique training approach allows \textbf{FetalCLIP} to effectively learn the intricate anatomical features present in fetal ultrasound images, resulting in robust representations that can be used for a variety of downstream applications.     In extensive benchmarking across a range of key fetal ultrasound applications, including classification, gestational age estimation, congenital heart defect (CHD) detection, and fetal structure segmentation, FetalCLIP outperformed all baselines while demonstrating remarkable generalizability and strong performance even with limited labeled data. We plan to release the FetalCLIP model publicly for the benefit of the broader scientific community.}




\maketitle

\section{Introduction}
Prenatal care is historically known to be transformed by the integration of ultrasound technology. Widely recognized for its accessibility, safety, and cost-effectiveness, ultrasound now plays a critical role in monitoring fetal development in real time and facilitating the early detection of congenital abnormalities \cite{whitworth10,Salomon2013,Salomon2011,Khalil2024}. However, despite its transformative role, fetal ultrasound image interpretation remains inherently subjective and heavily operator-dependent, often challenged by subtle visual cues, complex fetal anatomy, and significant inter-observer variability \cite{Chan2009,maraci2014}. These challenges can lead to inconsistencies in clinical assessments, particularly in resource-limited settings where access to highly trained sonographers is restricted \cite{Wanyonyi2017}. Studies have shown significant inter-observer variability in fetal ultrasound measurements, with differences of approximately ±4.9\% for head circumference (HC), ±8.8\% for abdominal circumference (AC), and ±11.1\% for femur length (FL) \cite{sarris2012intra}. These variabilities highlight the subjectivity and potential inconsistencies in fetal ultrasound assessments, emphasizing the need for AI-powered tools to enhance diagnostic objectivity and accuracy. These variabilities underscore the subjectivity and potential inconsistencies in fetal ultrasound assessments, highlighting the urgent need for robust AI-powered tools that enhance diagnostic objectivity, accuracy, and accessibility in fetal ultrasound imaging.

Recent advances in artificial intelligence (AI), particularly in foundation models, have demonstrated remarkable capabilities in improving medical imaging analysis \cite{azad2023,awais2025}. These models, pretrained on large-scale datasets, enable powerful feature extraction and knowledge transfer to downstream tasks, enhancing diagnostic precision, optimizing clinical workflows, and broadening access to expert-level interpretation.
However, existing foundation models \cite{clip, biomedclip, unimedclip, medclip, elixr, medfilip, flair, eyeclip, plip, conch, unipathology, echoclip, echoprime, echoapex, ctclip, fmct} exhibit fundamental limitations when applied to fetal ultrasound.
For instance, CLIP \cite{clip}, primarily trained on natural images, lacks the domain-specific anatomical knowledge necessary for medical imaging. BiomedCLIP \cite{biomedclip}, while tailored for the biomedical domain, is primarily optimized for text-based biomedical knowledge retrieval rather than complex image-text reasoning required for fetal ultrasound interpretation, limiting its ability to effectively capture fine-grained anatomical details. UniMed-CLIP \cite{unimedclip}, though leveraging a large-scale open-source medical multimodal dataset, demonstrates variable performance across different imaging modalities and remains largely unexplored in the context of fetal ultrasound.
Complementary to the aforementioned approaches, numerous efforts have been made to develop foundation models for specific medical imaging modalities, including chest X-ray \cite{medclip, elixr, medfilip}, ophthalmic imaging \cite{flair, eyeclip}, pathology \cite{plip, conch, unipathology}, echocardiography \cite{echoclip, echoprime, echoapex}, chest CT \cite{ctclip}, and head CT \cite{fmct}.
These models struggle to generalize to the unique challenges posed by fetal imaging, failing to capture subtle morphological variations that are crucial for assessing fetal development. Moreover, most AI solutions in fetal ultrasound rely on limited datasets and do not achieve the level of generalizability required for robust clinical deployment, particularly in detecting rare fetal conditions. Encoding the entirety of fetal anatomy while preserving diagnostically critical features remains a significant challenge for existing AI methods. To overcome these barriers, a dedicated fetal ultrasound-specific foundation model is required to harness the full potential of AI for advancing prenatal care.

Here, we introduce FetalCLIP, a novel visual-language foundation model explicitly engineered for fetal ultrasound analysis, trained on the most comprehensive dataset of its kind. FetalCLIP is pretrained at an unprecedented scale ({\color{red}Fig. \ref{fig:fetalclip_overview}a}), leveraging 207,943 fetal ultrasound images with corresponding GPT-4o-generated captions and 2,092 expert-annotated image-caption pairs from a fetal ultrasound textbook. This extensive dataset, which was curated by authors in collaboration with medical experts, covers a broadspectrum of fetal anatomical structures and developmental stages, ensuring both diversity and robustness. FetalCLIP incorporates an innovative multimodal contrastive learning strategy that integrates visual and textual representations of fetal ultrasound data, allowing the model to effectively align anatomical structures with diagnostic descriptions and enhance interpretability. This advanced pretraining paradigm empowers FetalCLIP to learn rich, generalizable representations of fetal ultrasound scans and effectively transfer this knowledge to a diverse range of downstream tasks, including zero-shot classification of standard fetal views, congenital heart disease (CHD) detection from ultrasound videos, segmentation of fetal anatomical structures, and feature extraction for downstream fetal ultrasound tasks, ensuring adaptability across clinical applications. Unlike existing models, FetalCLIP’s dual-modality learning approach allows it to discern subtle, clinically actionable patterns in fetal ultrasound images, surpassing the capabilities of vision-only models and yielding substantial improvements in diagnostic accuracy and clinical interpretability. 

By addressing the critical gaps in fetal ultrasound analysis, FetalCLIP represents a significant step toward more reliable, accessible, and AI-driven prenatal diagnostics. FetalCLIP undergoes rigorous evaluation across multiple downstream tasks to assess its generalizability and clinical applicability. It achieves an 87.1\% F1 score in zero-shot classification of standard fetal views, outperforming existing foundation models and a state-of-the-art open-source fetal view classifier trained with supervised learning \cite{sononet}. In the critical task of congenital heart disease (CHD) detection from ultrasound videos, FetalCLIP demonstrates 6.92\% improvement in AUC over previous models, showcasing its ability to detect subtle morphological variations crucial for early diagnosis. Furthermore, for fetal anatomical segmentation, it attains an average Dice Similarity Coefficient (DSC) of 84.22\% across three different fetal anatomical planes, highlighting its proficiency in delineating fetal structures with high precision. These findings firmly establish FetalCLIP as a pivotal advancement, bridging the gap between human-level expertise and AI-driven prenatal diagnostics and setting a new benchmark for the field of fetal ultrasound analysis.


\section{Results}

\subsection{Overview of FetalCLIP}

\begin{figure}[ht!]
    \centering
    \includegraphics[width=\linewidth]{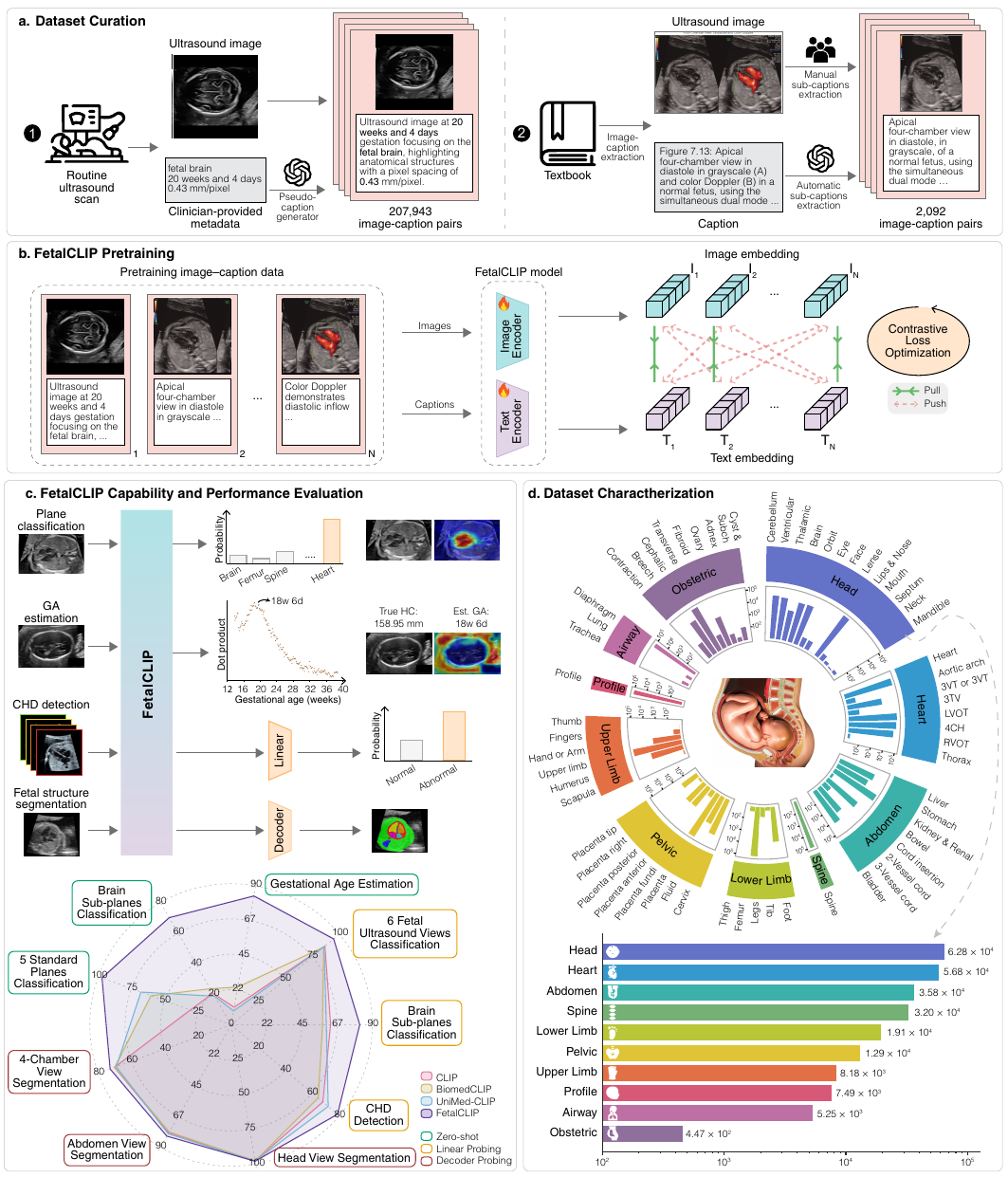}
    \caption{
    \textbf{Overview of FetalCLIP development and performance.}
    \textbf{a}, Dataset curation of fetal ultrasound image-caption pairs used for the FetalCLIP pretraining. The pretraining data was curated from two sources: (1) routine pregnancy ultrasound scans, comprising 207,943 images with corresponding LLM-generated pseudo captions, which incorporate clinicians' labels, gestational age, and pixel spacing; and (2) 2,092 image-caption pairs derived from a fetal ultrasound textbook.
    \textbf{b}, FetalCLIP pretraining step through contrastive learning, maximizing similarity between paired image-captions while minimizing similarity to unrelated pairs.
    \textbf{c}, Schematic diagram illustrating FetalCLIP’s capability and radar plot demonstrating FetalCLIP's superior performance over existing vision-language foundation models across diverse fetal ultrasound tasks, including fetal planes classification, congenital heart disease (CHD) detection, and fetal structures segmentation on different views.
    \textbf{d}, Distribution of routine pregnancy ultrasound scan data, which constitutes the largest portion of the FetalCLIP pretraining data.
    }
    \label{fig:fetalclip_overview}
\end{figure}

FetalCLIP is a vision-language foundation model for fetal ultrasound analysis, developed by leveraging a large-scale dataset of paired fetal ultrasound images and captions ({\color{red}Fig. \ref{fig:fetalclip_overview}a}). The dataset comprises 207,943 images from routine clinical scans encompassing 64 diverse clinician-labeled keywords across 6,493 patients with a mean gestational age of $148\pm16$ days (21 weeks 1 day ± 2 weeks 2 days), supplemented by 2,092 image-caption pairs derived from a fetal ultrasound textbook \cite{fhtextbook} primarily focused on the fetal heart.
As expert-provided image-level text descriptions were absent for the clinical dataset, we used GPT-4o to generate clinically sound captions based on gestational age, clinical labels, and pixel spacing, with details provided in the Methods. This approach ensured each image was paired with a unique, contextually relevant textual description. We adopted the contrastive language-image pretraining (CLIP) \cite{clip} framework to pretrain FetalCLIP by aligning images with their corresponding textual descriptions in a shared embedding space while minimizing similarity to unrelated pairs ({\color{red}Fig. \ref{fig:fetalclip_overview}b}). The FetalCLIP architecture includes a ViT-L \cite{vit} image encoder, a Byte-Pair Encoding tokenizer \cite{bpe}, and a text encoder \cite{radford2019language} capable of processing up to 117 tokens to accommodate the maximum token length in the rich clinical text descriptions.
Our extensive evaluations demonstrated the superior performance of FetalCLIP over existing vision-language foundation models across various tasks in fetal ultrasound analysis ({\color{red}Fig. \ref{fig:fetalclip_overview}c}), attributed to its pretraining on a large-scale dataset comprising predominantly clinical data with diverse fetal ultrasound keywords ({\color{red}Fig. \ref{fig:fetalclip_overview}d}).

\subsection{Zero-shot classification of standard fetal views}
\label{ssec:resultszeroshotclassification}

\begin{figure}[ht!]
    \centering
    \includegraphics[width=\linewidth]{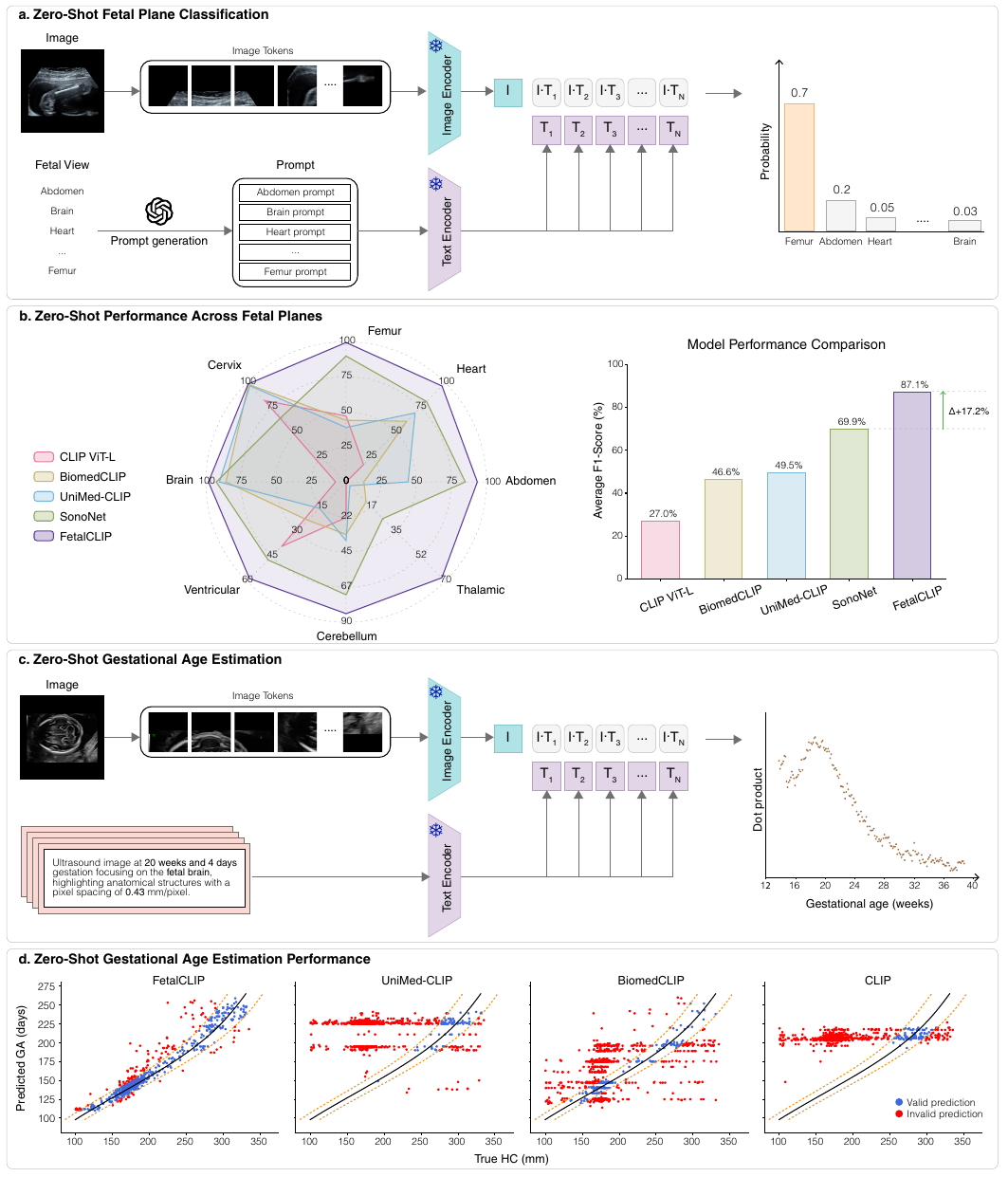}
    \caption{
    \textbf{Zero-shot capabilities of FetalCLIP.}
    \textbf{a}, Illustration of zero-shot fetal plane classification. We leveraged an LLM to generate prompts for a set of predefined candidate planes (detailed in {\color{red}Extended Data Fig. \ref{prompts_for_inference}}). The predicted plane was determined by identifying the highest similarity between the image embedding and prompt embeddings.
    \textbf{b}, Zero-shot performance in distinguishing five standard fetal planes and three brain subplanes. FetalCLIP achieved the highest accuracy with an average F1 score of 87.1\%, outperforming the specialist model SonoNet by 17.2\%.
    \textbf{c}, Illustration of zero-shot GA estimation. A similarity map was computed between the image embeddings and prompts embeddings spanning 14 to 40 weeks of GA. We then subsequently postprocessed the similarity map to predict GA.
    \textbf{d}, GA estimation performance of visual-language foundation models. The \textbf{\color{blue}blue} points represent valid predictions, while the \textbf{\color{red}red} points indicate invalid predictions. The \textbf{black} line represents the 50th percentile of the quantile regression population, and the \textbf{\color{orange}orange} lines represent the 2.5th and 97.5th percentiles of the population as provided by the WHO \cite{fetalcalculator}. Unlike FetalCLIP, other models demonstrated no ability to infer GA from fetal ultrasound head images.
    }
    \label{fig:zero_shot}
\end{figure}

We conducted a study to evaluate FetalCLIP's capability in classifying standard fetal ultrasound views using unseen data from different hospitals without any adaptation ({\color{red}Fig. \ref{fig:zero_shot}a}).
FetalCLIP was compared against notable models in the field: (1) SonoNet \cite{sononet}, a specialized model specifically trained for classifying standard views of fetal ultrasound; (2) CLIP \cite{clip}, a visual-language model (VLM) for natural images; and two visual-language foundation models tailored for the general medical domain, (3) BiomedCLIP \cite{biomedclip} and (4) UniMed-CLIP \cite{unimedclip}.
We conducted the evaluation using the Planes DB \cite{planes_db} dataset, which consists of fetal ultrasound images acquired from two hospitals. The models were employed to classify five anatomical planes—abdomen, brain, cervix, femur, and thorax—as well as three subplanes within the brain—transcerebellum, transthalamic, and transventricular.
Our experiment ({\color{red}Fig. \ref{fig:zero_shot}b}) demonstrated that FetalCLIP achieved an average F1 score of 87.1\%, outperforming SonoNet by a notable margin of 17.2\%, UniMed-CLIP by 37.6\%, BiomedCLIP by 40.5\%, and CLIP by 60.1\%.
The findings also highlight that CLIP lacks the contextual information required to differentiate between different fetal ultrasound anatomical views, and both BiomedCLIP and UniMed-CLIP struggle in distinguishing between brain subplanes.
Our confusion matrix analysis ({\color{red}Extended Data Fig. \ref{ext_fig_conf_matrix}}) further added that SonoNet struggled with identifying the cervix view, while UniMed-CLIP could not distinguish the spine view from other fetal planes.
In contrast, by harnessing large-scale fetal ultrasound data and incorporating semantic knowledge from textual descriptions during pretraining, FetalCLIP achieved remarkable accuracy in zero-shot classification of standard anatomical views.

\subsection{Zero-shot gestational age estimation}
The FetalCLIP pretraining stage requires the model to incorporate gestational age in order to precisely align fetal ultrasound images with their corresponding text descriptions. This enables FetalCLIP to estimate gestational age (GA) directly from the images ({\color{red}Fig. \ref{fig:zero_shot}c}) to a certain degree of precision, without additional fine-tuning, as detailed in the Methods section. To evaluate its zero-shot performance for this specific task, we leveraged the HC18 \cite{hc18} dataset, which includes fetal brain images along with head circumference (HC) annotations and pixel spacing. As the true GA was not available, we defined a prediction to be valid if the true HC falls within the 2.5th to 97.5th percentile range of HC for the predicted GA, which was calculated using the quantile regression method established by the WHO \cite{fetalcalculator}. FetalCLIP achieved a prediction validity rate of 83.5\%, while other models produced much lower validity rates (CLIP: 11\%, BiomedCLIP: 24\%, and UniMed-CLIP: 9\%).
This evaluation ({\color{red}Fig. \ref{fig:zero_shot}d}) also highlights that while existing visual-language foundation models failed to reliably infer GA from images, FetalCLIP effectively produced a high proportion of valid GA predictions.
Our further investigation demonstrated that FetalCLIP achieved a higher validity rate of 89\% within the range of true HC values associated with the 25th to 75th percentile of the pretraining GA distribution (20 weeks 0 days to 21 weeks 6 days). This suggests that FetalCLIP's pretraining data influenced its zero-shot performance in estimating GA.

\subsection{FetalCLIP as strong feature extractor for fetal US}
Motivated by the growing need for efficient tuning to adapt large pre-trained models to diverse applications—an essential consideration in building large and complex AI systems \cite{Lu2024,navid,videoglamm,glamm}—we assessed the capability of the FetalCLIP image encoder to extract generalizable features for downstream fetal ultrasound tasks. In this setup, the image encoder was entirely frozen, while a lightweight network was trained to utilize the extracted features for a specific ultimate task—e.g. a linear layer for classification. This setting heavily relies on the image encoder to provide robust image representations. We evaluated the image encoder on the following tasks: (1) standard view classification in fetal ultrasound, (2) congenital heart defect (CHD) detection from fetal ultrasound videos, and (3) segmentation of fetal anatomical structures across different views. Our experiments ({\color{red}Fig. \ref{fig:fetalclip_overview}c}) demonstrated that FetalCLIP consistently outperformed other visual-language foundation models across all tasks, highlighting its superior performance as a feature extractor capable of delivering generalizable image representations for fetal ultrasound analysis.

\subsubsection{Probing FetalCLIP for fetal views classification}
\begin{figure}[ht!]
    \centering
    \includegraphics[width=\linewidth]{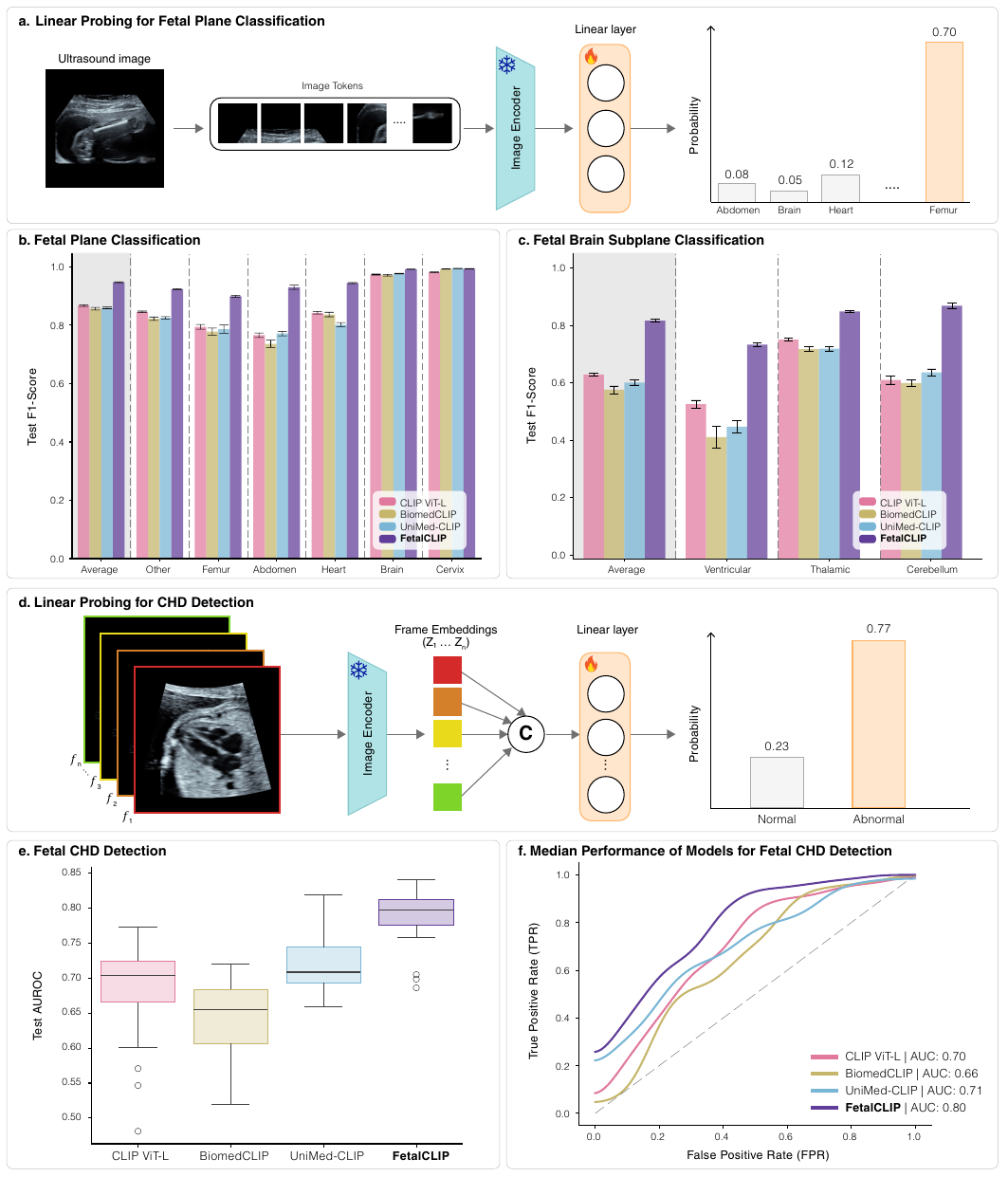}
    \caption{
    \textbf{Linear probing for classification tasks.}
    \textbf{a,} Schematic of linear probing for classifying different fetal planes. The image encoder of a visual-language foundation model was used to extract image embeddings, followed by a trainable linear layer for classification.
    \textbf{b-c,} F1 scores in the testing set for fetal plane and brain subplane classification, from 5-fold cross-validations with five different seeds. The bars represent the mean F1 scores, while the error bars indicate the standard deviation.
    \textbf{d,} Illustration of linear probing for CHD detection from an ultrasound clip. Embeddings were extracted from each image in the clip and concatenated. A trainable linear layer was then applied to leverage the combined embeddings for classification.
    \textbf{e,} AUROC comparisons for CHD detection across 5-fold cross-validations with 5 different seeds.
    \textbf{f,} ROC curve for CHD prediction showing the median performance of each model.
    }
    \label{fig:linear_probing}
\end{figure}

In some clinical practices, adapting a foundation model is essential to achieve optimal accuracy in identifying specific fetal views from datasets obtained from different hospitals. To this end, we investigated the FetalCLIP ability to provide robust features for accurately distinguishing six fetal ultrasound views and three brain subplanes using the Planes DB dataset. We attached a single linear layer to the frozen FetalCLIP image encoder, harnessing its image representations and transforming them into predictions for the set of output views ({\color{red}Fig. \ref{fig:linear_probing}a}). Across all views, FetalCLIP achieved significantly higher F1 scores compared to CLIP, BiomedCLIP, and UniMed-CLIP ({\color{red}Fig. \ref{fig:linear_probing}b-c}). In addition, we also evaluated their performance in a data-efficient setting, using data from only a few patients for training. Similarly, FetalCLIP consistently outperformed both the natural and medical vision foundation models ({\color{red}Extended Data Fig. \ref{fig_few_patients}a-b}). Using data from only 32 patients, FetalCLIP demonstrated comparable or even superior accuracy to UniMed-CLIP trained on the full dataset of 717 patients.

\subsubsection{Probing FetalCLIP for video-based CHD detection}
Clinicians are often required to analyze ultrasound videos to assess fetal conditions, such as detecting abnormalities in the fetal heart. However, developing AI solutions for such tasks is challenging given the limited availability of annotated ultrasound video data. Thus, leveraging pretrained models is critical for enhancing model generalizability. Motivated by this, we adapted FetalCLIP to analyze fetal ultrasound videos focusing on the 4-chamber view, with the aim of distinguishing between normal fetal hearts and those affected by congenital heart disease (CHD) ({\color{red}Fig. \ref{fig:linear_probing}d}). We leveraged FetalCLIP to extract image features from each frame of the ultrasound videos. These frame-level features were then combined (see Methods), and a linear layer was applied to classify the fetal heart as either normal or having CHD. Our experiment demonstrated that FetalCLIP outperformed the other foundation models by a substantial margin ({\color{red}Fig. \ref{fig:linear_probing}e-f}), achieving a mean AUROC of 78.72\%, whereas CLIP, BiomedCLIP, and UniMed-CLIP achieved 67.88\% ($P<10^{-5}$), 64.32\% ($P<10^{-6}$), and 71.8\% ($P<10^{-3}$) respectively. This showcases the excellent adaptability of FetalCLIP in analyzing fetal ultrasound videos.

\subsubsection{Probing FetalCLIP for segmenting fetal structures}
\begin{figure}[ht!]
    \centering
    \includegraphics[width=\linewidth]{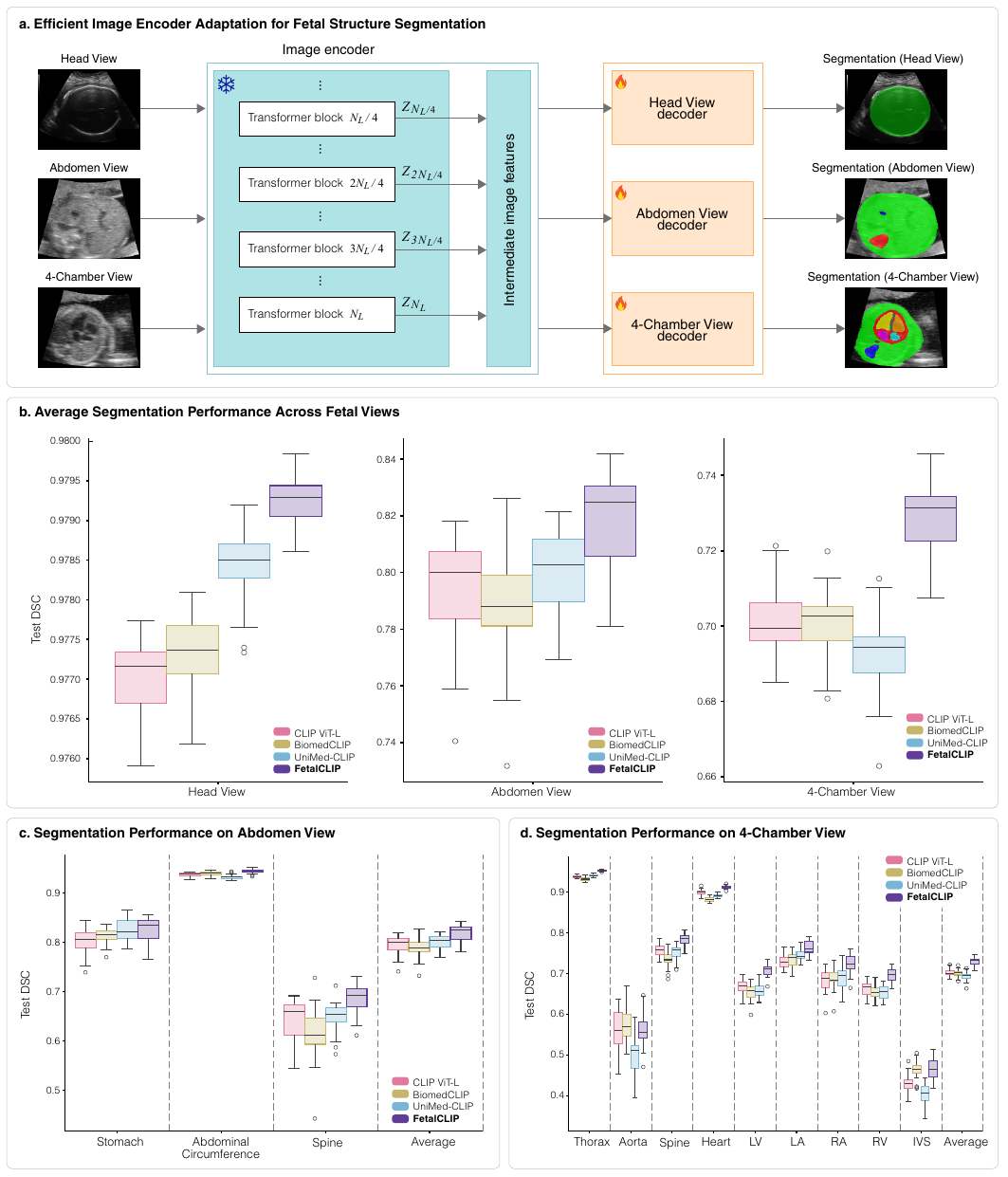}
    \caption{
        \textbf{Segmentation of various fetal structures across different views.}
        \textbf{a,} Illustration of the efficient adaptation of an image encoder for segmenting fetal structures in different views. A lightweight decoder was developed to leverage intermediate embeddings from the image encoder for segmentation. $N_L$ denotes the number of transformer blocks in the image encoder, which is 12 for ViT-B and 24 for ViT-L, respectively.
        \textbf{b,} Average segmentation performance across structures within each view (head, abdomen, and 4-chamber) evaluated over 5-fold cross-validations with five different seeds.
        \textbf{c-d,} Dice Similarity Coefficient (DSC) for individual structures in the abdomen view and 4-chamber view, respectively. LV, left ventricle; LA, left atrium; RA, right atrium; RV, right ventricle; IVS, interventricular septum.
    }
    \label{fig:decoder_probing}
\end{figure}

Accurate pixel-level classification is critical for precise growing fetal biometry calculations \cite{hc18,Slimani2023}. We investigated the foundation models' ability to provide fine-grained intermediate image features essential for localizing fetal anatomical structures. We apply a lightweight decoder with few parameters ($\sim$1.3 million parameters for ViT-B and $\sim$1.6 million for ViT-L encoders) to utilize the intermediate image features for accurate segmentation of fetal anatomical structures ({\color{red}Fig. \ref{fig:decoder_probing}a}). We conducted segmentation experiments on three fetal views to segment various structures: 1) brain view (head), 2) abdomen view (abdomen, stomach, and spine), and 3) 4-chamber view (nine structures, see {\color{red}Extended Data Table \ref{tab:ext_tab_downstream_setting}}). We reported the average Dice Similarity Coefficient (DSC) for each view and weighted each structure equally. FetalCLIP achieved DSC of 97.92\%, 81.82\%, and 72.91\% for brain, abdomen, and 4-chamber views, respectively, surpassing UniMed-CLIP by 0.08\% ($P<10^{-6}$), 1.7\% ($P<10^{-3}$), and 3.64\% ($P<10^{-7}$) ({\color{red}Fig. \ref{fig:decoder_probing}b-d}). A similar trend was observed in data-efficient settings ({\color{red}Extended Data Fig. \ref{fig_few_patients}c-e}), where FetalCLIP consistently outperformed other foundation models. These results demonstrated that FetalCLIP can serve as a robust feature extractor for fetal ultrasound tasks requiring fine-grained anatomical segmentation.

\subsection{FetalCLIP interpretability}
\begin{figure}[ht!]
    \centering
    \includegraphics[width=\linewidth]{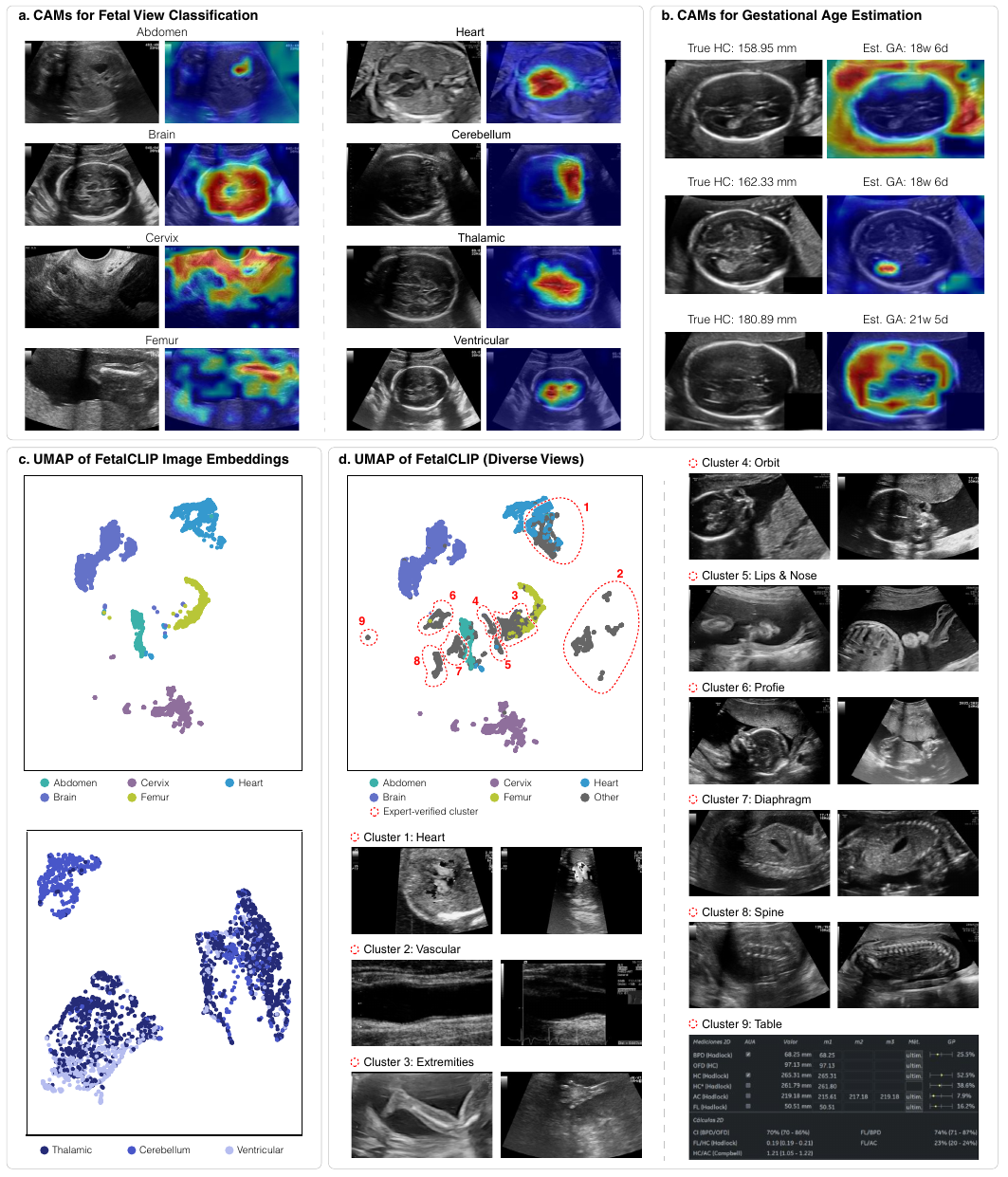}
    \caption{
        \textbf{FetalCLIP interpretation studies.}
        \textbf{a,} Class activation maps (CAMs) of FetalCLIP highlighting important regions for predicting fetal ultrasound views. FetalCLIP effectively emphasized important structures when determining anatomical views.
        \textbf{b,} CAMs of FetalCLIP when estimating gestational age, highlighting brain structures and surrounding head circumference regions.
        \textbf{c,} UMAP visualization of FetalCLIP image embeddings, corresponding to five standard fetal planes and three brain subviews.
        \textbf{d,} Image embeddings of diverse fetal views. Images containing similar information were mapped into close proximity. We manually inspected several points in each cluster belonging to the "Other" class to determine the dominant views within each cluster, which were then verified by clinicians.
    }
    \label{fig:interpretation_studies}
\end{figure}

To analyze the FetalCLIP reliability from the clinical perspective, we conducted interpretation studies to investigate how the model derives its prediction and assess whether it aligns with clinical practice. We employed class activation mapping (CAM) via ScoreCAM \cite{scorecam} to visualize the importance of image regions to FetalCLIP's decisions, as depicted in {\color{red}Fig. \ref{fig:interpretation_studies}a-b}. The CAMs in {\color{red}Fig. \ref{fig:interpretation_studies}a} suggested that FetalCLIP can effectively highlight key fetal landmarks when identifying anatomical views, such as stomach in the abdomen view, femur bone, heart circumference, cerebellar hemispheres, and cavum septum pellucidi. This showcased FetalCLIP's ability to localize specific structures within fetal ultrasound images. Furthermore, as visualized in {\color{red}Fig. \ref{fig:interpretation_studies}b}, FetalCLIP highlighted regions surrounding the head circumference and some brain structures, such as the choroid plexus, to estimate gestational age.

To gain a deeper understanding of FetalCLIP's image representations, we utilized the Uniform Manifold Approximation and Projection (UMAP) \cite{umap} to visualize FetalCLIP image embeddings in a two-dimensional space. We found that FetalCLIP could effectively cluster five standard fetal planes ({\color{red}Fig. \ref{fig:interpretation_studies}c}) and differentiate between brain subviews ({\color{red}Fig. \ref{fig:interpretation_studies}d}).
Our further investigation ({\color{red}Fig. \ref{fig:interpretation_studies}e}) revealed that FetalCLIP could cluster other fetal views such as profile and spine, as well as non-anatomical elements like tables, and could map images containing related anatomical structures into close proximity, such as the fetal extremities, which includes the leg, feet, arm, and hand. These observations were then verified by clinicians to confirm their validity.
Thus, this finding showcases FetalCLIP’s ability to enable automatic plane classification and clustering, potentially improving workflow efficiency in clinical practice.


\section{Discussion}

The advent of foundation models has shifted the field of image analysis toward an exciting paradigm.
However, unlike in the natural domain, developing foundation models for the medical domain is more challenging due to the high heterogeneity of medical modalities. While recent studies have shown that modality-specific medical foundation models often outperform general medical foundation models \cite{plip, conch, flair, eyeclip, unimedclip, medfilip}, both approaches have largely ignored fetal ultrasound. This slow advancement in fetal ultrasound can be attributed to the data scarcity in this domain that is both sufficient in quantity and contextual information.
This study introduces FetalCLIP as the first visual-language foundation model designed explicitly for fetal ultrasound.

In this study, our findings suggest that despite the absence of large image-clinical text description pairs, a strong visual-language foundation model can be developed by leveraging routine pregnancy scans with limited contextual information, supplemented by image-caption pairs from a textbook. Unlike other foundation models that place limited emphasis on fetal ultrasound, FetalCLIP is highly adaptable across various fetal ultrasound tasks. Despite not being explicitly trained for specific tasks, FetalCLIP achieved excellent zero-shot classification of different fetal anatomical planes, surpassing a specialist model for view classification (SonoNet \cite{sononet}), even when tested on unseen data from multiple hospitals. Pretrained with over 64 clinician-labeled keywords, FetalCLIP could reduce the labor-intensive process of manually identifying fetal ultrasound images, thus improving the accuracy and efficiency of prenatal assessment, especially in rural or low-resource environments. Furthermore, FetalCLIP is equipped with zero-shot capability to estimate gestational age from fetal images with remarkable accuracy. This highlights the presence of the FetalCLIP knowledge to extract meaningful information from fetal structures for assessing fetal growth. However, FetalCLIP struggles to accurately estimate gestational age for fetuses in early and late gestation. We hypothesize that this limitation arises from the distribution of its pretraining data, with the majority of the images acquired during the second trimester. Expanding pretraining data to cover a broader gestational age range could further improve performance.

This study also demonstrates that FetalCLIP can serve as a robust feature extractor. Our extensive downstream experiments across diverse fetal ultrasound tasks revealed substantial performance gains of FetalCLIP over other foundation models. The experiments included standard fetal plane classification, brain subviews classification, CHD detection from four-chamber videos, and segmentation of different fetal structures in the head, abdomen, and four-chamber views. These results establish our visual-language foundation model as the most preferred pretrained model for developing AI models to solve challenging problems in fetal ultrasound analysis, especially in data-efficient settings. In addition, these findings align with recent studies reporting the superiority of modality-specific foundation models over their general medical counterparts when applied to their respective targeted modalities \cite{plip,conch,flair,eyeclip,unimedclip,medfilip}.

In addition to FetalCLIP's superior performance compared to other visual-language foundation models, our interpretability analysis underscores FetalCLIP's reliability and alignment with clinical practice. Using ScoreCAM, we demonstrated that FetalCLIP effectively localizes relevant anatomical structures and regions of interest when identifying anatomical views ({\color{red}Fig. \ref{fig:interpretation_studies}a}), such as the stomach in the abdomen view and cerebellar hemispheres in the transcerebellar plane. This CAM analysis also demonstrates that, when estimating gestational age, the model focuses on regions such as the head circumference and brain structures, including the choroid plexus ({\color{red}Fig. \ref{fig:interpretation_studies}b}). Moreover, UMAP visualizations ({\color{red}Fig. \ref{fig:interpretation_studies}c-e}) revealed that FetalCLIP's embeddings can effectively cluster standard fetal planes and differentiate between brain subviews, while mapping related anatomical structures into close proximity, e.g. fetal extremities.


FetalCLIP's significant gains in zero-shot, transfer learning, and interpretability contribute to advancing fetal ultrasound image analysis, with future improvements possible through enriched pretraining data.
While our zero-shot results demonstrate substantial performance gains of FetalCLIP over existing visual-language foundation models, we anticipate that the FetalCLIP's zero-shot capability for detecting abnormalities in fetal ultrasound scans may still be limited. This limitation likely arises from the fact that the majority of FetalCLIP pretraining data was collected from routine pregnancy scans in the second trimester which lacked information on fetal health conditions. Nevertheless, FetalCLIP exhibits superior performance and higher adaptability across various tasks compared to existing foundation models, which were typically trained on millions of image-caption pairs. This leaves immense potential for future studies to enhance FetalCLIP's representations by incorporating more diverse fetal ultrasound images and richer image-level descriptions. Additionally, due to computational constraints and to facilitate a fair benchmark with compared foundation models, we restricted our experiments to $224\times224$ image size. Notably, leveraging higher resolutions could elevate FetalCLIP's capabilities by providing clear fine-grained details such as valves in fetal hearts and brain structures.
This study also underscores potential future exploration that includes expanding the pretraining data to cover wider data distribution, more diverse clinical scenarios, and a broader range of image types, as well as extending FetalCLIP to a video encoder for better capturing spatio-temporal features, thus broadening its applicability and impact in fetal ultrasound analysis.
By releasing FetalCLIP to the public, we aim to foster the advancement of fetal ultrasound analysis by enabling researchers and clinicians to develop innovative applications on top of this strong foundation model.



\section{Methods}

\subsection{FetalCLIP pretraining data curation}

FetalCLIP was pretrained using two data sources, curated to develop a robust and generalizable visual-language foundation model for fetal ultrasound image analysis. The first data source consists of 207,943 fetal ultrasound images from Corniche Hospital Abu Dhabi, a tertiary-level referral unit providing expert care to women and neonates. This data was supplemented by 2,092 image-caption pairs from a fetal ultrasound textbook \cite{fhtextbook} emphasizing fetal heart conditions, providing critically important clinical information for fetal health assessment in clinical practice.

\subsubsection{Private hospital dataset}

This data source constitutes the largest portion of the FetalCLIP pretraining dataset. The images were obtained from routine pregnancy scans at the hospital to monitor fetal development, often performed during the second trimester. During this stage, the fetus begins to resemble a baby, and sonographers examine various anatomical planes to detect any potential health abnormalities. The average gestational age in this data source is $148 \pm 16$ days, with 50\% of the images lying between 20 weeks 0 days and 21 weeks 6 days. Some images were accompanied by clinicians' text annotations embedded within the images, from which we identified 64 distinct keywords predominantly related to fetal anatomical structures ({\color{red}Fig. \ref{fig:fetalclip_overview}d}). However, no associated medical report is available for each image. For this study, we utilized B-mode images and leveraged text annotations written by clinicians to serve as the basis for image labeling and text description generation for each image.

A considerable effort was made to preprocess, clean, and standardize this data source. We began by detecting clinicians' text annotations using the EasyOCR \cite{easyocr} library, followed by manual text processing to clean and standardize the detected annotations, resulting in the 64 keywords. In addition, to prevent potential model shortcuts, such as reliance on text annotations within the images, we designed a robust image preprocessing pipeline. The ultrasound fan region was extracted by first isolating the foreground, retaining all non-zero pixel values. We then applied the findContours function from the OpenCV \cite{opencv} library and identified the largest contour as the ultrasound fan region. To address text annotations, colored regions within the images were detected and inpainted using the Fast Marching Method \cite{inpainting_fmm}. These preprocessing steps removed confounding information from fetal ultrasound images, facilitating unbiased model training.

Based on the clinician annotations, we categorized this data source into three subgroups: (1) common standard anatomical views, (2) images with diverse clinical keywords, and (3) images without text annotations.

\noindent \textbf{Subgroup 1.} This subgroup comprised 88,045 images labeled across 12 standard anatomical views \cite{fusc}, including the abdomen, brain, cord, diaphragm, feet, femur, heart, kidney, lips \& nose, orbit, profile, and spine, with some examples shown in {\color{red}Extended Data Fig. \ref{ext_fig_corniche_dataset}a}. To ensure high-quality labeling, a confident learning algorithm \cite{northcutt2021confident} was employed to detect potentially mislabeled samples. This process involved training a 5-fold cross-validation (CV) model to classify the 12 views, using a patient-wise dataset split to avoid data leakage across folds. The algorithm identified mislabeled samples by comparing the model's confidence in its predictions with the provided labels. A total of 984 samples were identified as potentially mislabeled (examples shown in {\color{red}Extended Data Fig. \ref{ext_fig_corniche_dataset}b}) and subsequently excluded, resulting in a final dataset of 87,061 labeled images.

For the heart images, clinicians provided subview annotations for LVOT, RVOT, 4-CH, and 3VV/3VT, providing fine-grained information. In contrast, 468 brain images (out of 5,297) were labeled by an expert into three brain subviews, i.e. transthalamic, transcerebellum, and transventricular. These expert-labeled brain images were then used to train a 5-fold CV brain subview classifier. Pseudo-labeling was applied to the remaining brain images, assigning subviews to images with probabilities exceeding 90\%. Images with lower probabilities were categorized under the general "Brain" class.

\noindent \textbf{Subgroup 2.} This sub-data contains 73,972 images with diverse clinical labels, including cases where multiple anatomical structures were visible in a single frame. As illustrated in {\color{red}Extended Data Fig. \ref{ext_fig_corniche_dataset}c}, these images showcase the complexity of capturing multiple information within a single ultrasound frame, providing valuable examples for foundation model pretraining.

\noindent \textbf{Subgroup 3.} This subgroup consists of 79,757 unlabeled images. A 5-fold CV 12-view classifier trained on the first subgroup data was used to generate pseudo labels. Only images with classification probabilities exceeding 90\% were retained, resulting in 46,910 pseudo-labeled images. Furthermore, labeled heart subviews and brain subviews were used to train a 5-fold CV classifier for providing fine-grained information for pseudo-labels. Similarly, we assigned brain and heart images to their respective subviews if the model confidence exceeded 90\%.

\noindent \textbf{Caption generation.}
We generated captions for fetal ultrasound images by integrating information from the clinician text annotations or pseudo-labels, along with gestational age and pixel spacing. Instead of generating captions for each individual image, we efficiently constructed five caption templates for each unique set of clinician text annotations using GPT-4o, as illustrated in {\color{red}Extended Data Fig. \ref{ext_fig_prompts_training}a-b}. These multiple caption templates also acted as a form of text augmentation that enabled the model to learn more robust visual-language representations. The variation in gestational age and pixel spacing across images ensured that the five captions for each image were distinguishable from those of other ultrasound images.

\subsubsection{Fetal ultrasound textbook dataset}
We extracted 849 image-caption pairs from a textbook \cite{fhtextbook} focused on fetal heart conditions. Since most figures in the textbook contained multiple subfigures, we manually separated these subfigures into 2,092 independent images. The corresponding captions were then divided into subcaptions and refined using GPT-4o to ensure they were self-contained and to eliminate references to visual markers (e.g. arrows), as illustrated in {\color{red}Extended Data Fig. \ref{ext_fig_prompts_training}c}. Given that this data source was approximately 100 times smaller than the hospital data source, we upsampled the image-caption pairs from this source by a factor of 10 to increase its significance during FetalCLIP pretraining. In addition, we employed data sharding \cite{ilharco_gabriel_2021_5143773} where each shard contained unique image-caption pairs without duplicates. This strategy ensured that an image-caption pair from this data source did not appear multiple times in a single batch, which could otherwise negatively affect the contrastive learning process.

\subsection{FetalCLIP architecture and pretraining}
We developed the FetalCLIP model by adapting the architecture and training methodology employed in CLIP \cite{clip}. Additionally, our FetalCLIP pretraining code was built on top of the OpenCLIP \cite{ilharco_gabriel_2021_5143773,cherti2023reproducible} repository with some modifications to suit our objectives. The FetalCLIP image encoder utilizes a ViT-L \cite{vit} architecture, featuring an image input size of 224$\times$224, a patch size of 14$\times$14, and 24 transformer layers. For the text encoder, we implemented a text transformer \cite{radford2019language} with 12 transformer layers and a maximum input token of 117—40 tokens more than the original CLIP model—to accommodate clinical text descriptions, which are typically more detailed than natural captions. Both the image and text encoders project their respective inputs into a shared 768-dimensional space. We pretrained the FetalCLIP model to maximize the similarity between embeddings of paired fetal ultrasound image-caption data while minimizing the similarity of unpaired images and captions. This pretraining strategy enables FetalCLIP to derive semantically rich feature embeddings from fetal ultrasound images and their associated captions.

We applied data augmentation techniques during FetalCLIP pretraining, including random rotation ($\theta_{rotation} \in [-7^\circ, 7^\circ]$), translation ($\theta_{translation} \in [-0.05, 0.05]$), and color jittering ($\theta_{brightness}, \theta_{contrast}, \theta_{saturation} \in [0.85, 1.15]$). We pretrained FetalCLIP for 20 epochs using a learning rate of 5e-6, a warmup phase of 2,000 steps, a cosine scheduler, and a weight decay of 0.1. Mixed-precision training was implemented on 4$\times$RTX A6000 GPUs, allowing a batch size of 140 per GPU. Model checkpoints were saved after each epoch and evaluated on zero-shot standard view classification ({\color{red}as in Section \ref{ssec:resultszeroshotclassification} and Section \ref{ssec:methodzeroshotclassification}}), retaining the checkpoint with the highest average F1 score.
We explored various model initializations and found that fine-tuning the original CLIP in the general medical domain provided a strong initialization (accessed from \cite{pmc_vit_l_14}) for our FetalCLIP model, improving the zero-shot view classification performance from 85.2\% to 87.1\%. This underscores the importance of aligning a foundation model for natural domain to the medical domain as a crucial step for better initialization, resulting in improved performance on more specific modalities.

\subsection{Zero-shot view classification}
\label{ssec:methodzeroshotclassification}

For zero-shot standard view classification with vision-language foundation models (VLMs), we first defined target classes and provided five text prompts per class to enable prompt ensembling for improved robustness, as VLMs are sensitive to text prompts. We used GPT-4o to generate the text prompts for inferencing as detailed in {\color{red}Extended Data Fig. \ref{prompts_for_inference}}.
Our investigation revealed that specifically for FetalCLIP, incorporating caption templates of our routine prenatal scan data to generate text prompts for inference improved the average F1 score by 2.74\% across five standard fetal planes: \textit{abdomen}, \textit{brain}, \textit{femur}, \textit{heart}, and \textit{cervix}.
During testing, each text prompt was converted into a 768-dimensional text embedding using the VLM tokenizer and encoder, and embeddings were averaged across prompts for every class. Input images were similarly encoded into 768-dimensional embeddings, and classification was performed by selecting the class with the highest cosine similarity between the image embedding and the text embeddings. The average F1 score across all target classes was reported as the main metric for performance.

This zero-shot performance was evaluated using the Planes DB~\cite{planes_db} dataset. The dataset contains six classes: abdomen, femur, brain, thorax, cervix, and a class labeled "other," representing diverse additional views. The brain category is further divided into three fine-grained subviews: transcerebellum, transthalamic, and transventricular. Each image was preprocessed by symmetrically padding with zeros to form a square and subsequently resizing to a uniform dimension of 224$\times$224. We first evaluated the performance on distinguishing five standard views, excluding the "other" class, resulting in 8187 test images.
To ensure a fair comparison with the specialist model (SonoNet), nine standard views were selected as target classes, i.e. \textit{abdomen, brain, femur, heart, kidney, lips \& nose, profile, spine,} and \textit{cervix}.
Secondly, we evaluated the model's ability to classify the three brain subviews, resulting in 2949 test images. As SonoNet lacks dedicated classes for both cervix and transthalamic, we mapped the "other" class in SonoNet to cervix and transthalamic for the first and second evaluations, respectively.
In addition, the performance of the SonoNet models varied across sizes, with SN16, SN32, and SN64 achieving F1 scores of 67.5\%, 69.9\%, and 67.2\%, respectively, where the mid-sized model (SN32) demonstrated the best performance.

\subsection{Zero-shot gestational age estimation}
We utilized the HC18 dataset~\cite{hc18} to evaluate VLM's ability to estimate gestational age (GA). The dataset includes fetal brain images alongside head circumference (HC) measurements and pixel spacing information. Since the true GA values are not provided, our evaluation was based on the ground-truth HC. The relationship between HC and GA can be explained using the following quantile regression formula established by the WHO \cite{fetalcalculator}:
\begin{equation}
    \label{eq:ga_to_hc}
    HC = b_0 + b_1 t + b_2 t^2 + b_3 t^3 + b_4 t^4
\end{equation}
Here, $t$ represents GA in days, and $b_0, b_1, b_2, b_3, b_4 \in \mathbb{R}$ are coefficients that depend on the quantile of interest. Equation \ref{eq:ga_to_hc} applies to GA ranging from 14 to 40 weeks. Consequently, we restricted the analysis to images with HC values between 100 mm and 342 mm, corresponding to the 50th percentile HC at 14 and 40 weeks of GA, respectively, resulting in a test set of 814 brain images. To assess the plausibility of the predicted GA, we formulated a proxy validation task: predictions were deemed valid if the true HC fell within the 2.5th to 97.5th percentile range (95\% of the population distribution) associated with the predicted GA.

Inspired by \cite{echoclip}, we estimated GA by first extracting image embeddings and subsequently constructing text prompts that describe the brain view, pixel spacing, and GA values ranging from 14 weeks 0 days to 40 weeks. Similar to our zero-shot view classifier, we provided five text prompts for each GA to enable ensembling. The cosine similarity between the image and the text prompts was computed for each GA, and ensembling was performed by averaging the cosine similarities across the five prompts for a given GA. As a postprocessing step, the final GA prediction was selected as the median of the GAs corresponding to the top 15 text prompts with the highest cosine similarity to the image embeddings.
This postprocessing step enhanced the accuracy of GA estimation, raising the validity rate by +3.08\% compared to selecting the GA based solely on the highest cosine similarity.

\subsection{Probing FetalCLIP for downstream tasks}
\subsubsection{Experimental setup}
The probing experiments were conducted on downstream fetal ultrasound tasks to evaluate the quality of image embeddings from FetalCLIP compared to other foundation models. In these experiments, the image encoder was frozen, and a trainable prediction head
was attached. We compared FetalCLIP with both natural and medical vision-language foundation models across various fetal ultrasound benchmarks, with $P$ values computed using the Wilcoxon signed-rank test. Unless otherwise specified, for each evaluation, the corresponding dataset was split by patients into training (80\%) and testing (20\%) sets to prevent information leakage in the test data due to patient attributes, with stratified splitting employed for classification tasks. The prediction head was tuned and validated using the training set, leaving the testing set remain untouched during model development. In addition, we applied the same preprocessing steps as in the zero-shot experiments, resulting in standardized images with dimensions of 224×224. We then augmented the training images offline multiple times and stored them locally, enabling pre-computation of image embeddings to accelerate the training process. Details of the dataset used for downstream tasks along with the training configurations are presented in {\color{red}Extended Data Table \ref{tab:ext_tab_downstream_setting}}.

We considered two training scenarios: full-data training and data-efficient training with few patients. For full-data training experiments, we performed stratified 5-fold cross-validation on the training set. The prediction head was trained, and the model with the lowest validation loss was evaluated on the test set. To achieve reliable statistical evaluation, each fold was run five times with different random seeds, resulting in a total of 25 runs.
For data-efficient training, we randomly selected five support sets, each consisting of $N$ patients for training and $N$ for validation, while evaluation was performed on the testing set. Each support set was run with five different random seeds, resulting in a total of 25 outcomes per experiment.

\subsubsection{View classification}
The probing for view classification experiments was carried out using the Planes DB~\cite{planes_db} dataset, with the average F1 score computed as the performance metric. The dataset comprises six categories of fetal ultrasound views with three brain subviews. To harness VLM image embeddings, we attached a single trainable linear layer to the output of the VLM image encoder, projecting the embeddings for the view prediction task, which yields six output classes for view classification and three output classes for fine-grained brain subview classification.
We adhered to the original train-test split provided in the dataset \cite{planes_db}, resulting in 7129-5271 samples and 1543-1406 samples for the first and second view classification tasks, respectively.

\subsubsection{CHD detection}
We evaluated the transferability of image embeddings from foundation models to fetal ultrasound video analysis using an internal dataset. We collected 418 four-chamber fetal heart ultrasound videos, comprising 161 normal and 257 abnormal scans, with temporal lengths ranging from 16 to 128 frames, labeled by an expert in fetal cardiology to ensure diagnostic accuracy. To simplify the task, instead of processing the entire sequences, we extracted 16-frame clips for diagnosis. For videos with up to 64 frames, we sampled clips with approximately uniform spacing, while for longer videos, clips were sampled with a temporal stride of 4. This approach ensured that each clip spanned at least 50\% of the original video length, thus retaining adequate information to enable diagnosis. Frame-level image features were extracted and then combined to generate clip-level representations. We investigated two trivial feature combination methods, with averaging and concatenating achieving AUROCs of 74.66 ($P<10^{-2}$) and 78.72, respectively. To classify the heart as normal or abnormal, we implemented a single linear layer that operated on the clip-level representations. We reported the area under the receiver operating characteristic curve (AUROC) as the performance metric.

\subsubsection{Segmentation}

A lightweight decoder was developed to harness intermediate image representations from foundation models for segmenting fetal anatomical structures.
Specifically, we adapted the UNETR \cite{unetr} decoder for 2D applications while reducing its computational complexity and number of parameters.
The deconvolution layers were replaced with depthwise deconvolution layers, and convolutional blocks were replaced with depthwise separable convolution \cite{mobilenetv1} blocks with a kernel size of 3. This design modification resulted in a lightweight decoder with 1.32 M and 1.59 M parameters for the ViT-B-based and ViT-L-based encoders, respectively. We reported the Dice Similarity Coefficient (DSC) to assess the quality of the segmentation results. For views containing multiple segmented structures, we employed a multilabel segmentation approach, wherein a single pixel could belong to multiple structures (e.g., a pixel might simultaneously represent the four-chamber view, heart, and thorax). This approach required $N_s$ output channels for $N_s$ structures, with the average DSC across all structures computed to give equal importance to each structure.
Details of the data and hyperparameters for segmentation in brain, four-chamber, and abdominal views are provided in {\color{red}Extended Data Table \ref{tab:ext_tab_downstream_setting}}.

\section{Data Availability}
This study involved two publicly available datasets for evaluating models on a subset of tasks, as well as private datasets for pretraining and additional evaluations. The public datasets were obtained from PlanesDB (\href{https://zenodo.org/records/3904280}{https://zenodo.org/records/3904280}) and HC18 (\href{https://zenodo.org/records/1327317}{https://zenodo.org/records/1327317}). Due to regulations and privacy constraints, the private datasets cannot be released to the public.

\section{Code Availability}
The FetalCLIP code, including pretrained weights and prompts for inference, can be accessed at \href{https://github.com/BioMedIA-MBZUAI/FetalCLIP}{https://github.com/BioMedIA-MBZUAI/FetalCLIP}.

\section{Acknowledgements}
We gratefully acknowledge that this work was supported by GE Healthcare. We express our gratitude to Corniche Hospital in Abu Dhabi for providing prenatal scan data along with fetal heart scans, and to the Department of Health (DOH) Abu Dhabi for their support in approving the study which facilitates access to the anonymous data for internal purposes. We thank Alfred Z. Abuhamad for allowing us to leverage his book for foundation model pretraining. We also thank GE Healthcare for providing data and annotations used for downstream segmentation tasks in the 4-chamber and abdomen views.

\newpage
\bibliography{sn-bibliography}

\newpage




\newpage

\newpage

\newpage

\newpage

\newpage

\newpage
\begin{extended_figure}[ht!]
    \centering
    \includegraphics[width=\linewidth]{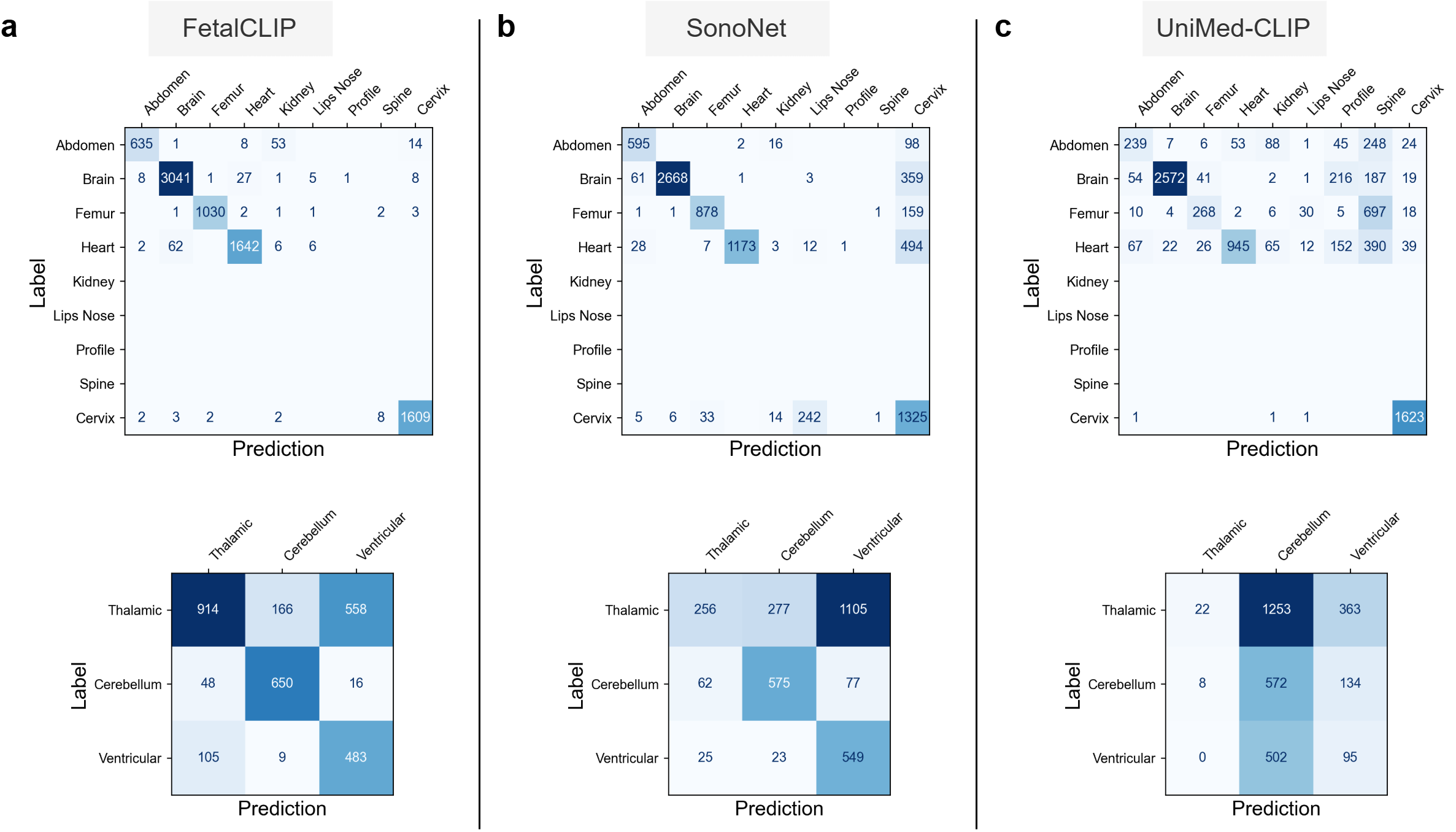}
    \caption{
        \textbf{Confusion matrices for five standard fetal planes and three brain subviews classifications.}
        \textbf{a-c,} represent the confusion matrices of FetalCLIP, SonoNet, and UniMed-CLIP, respectively.
    }
    \label{ext_fig_conf_matrix}
\end{extended_figure}

\newpage
\begin{extended_figure}[ht!]
    \centering
    \includegraphics[width=\linewidth]{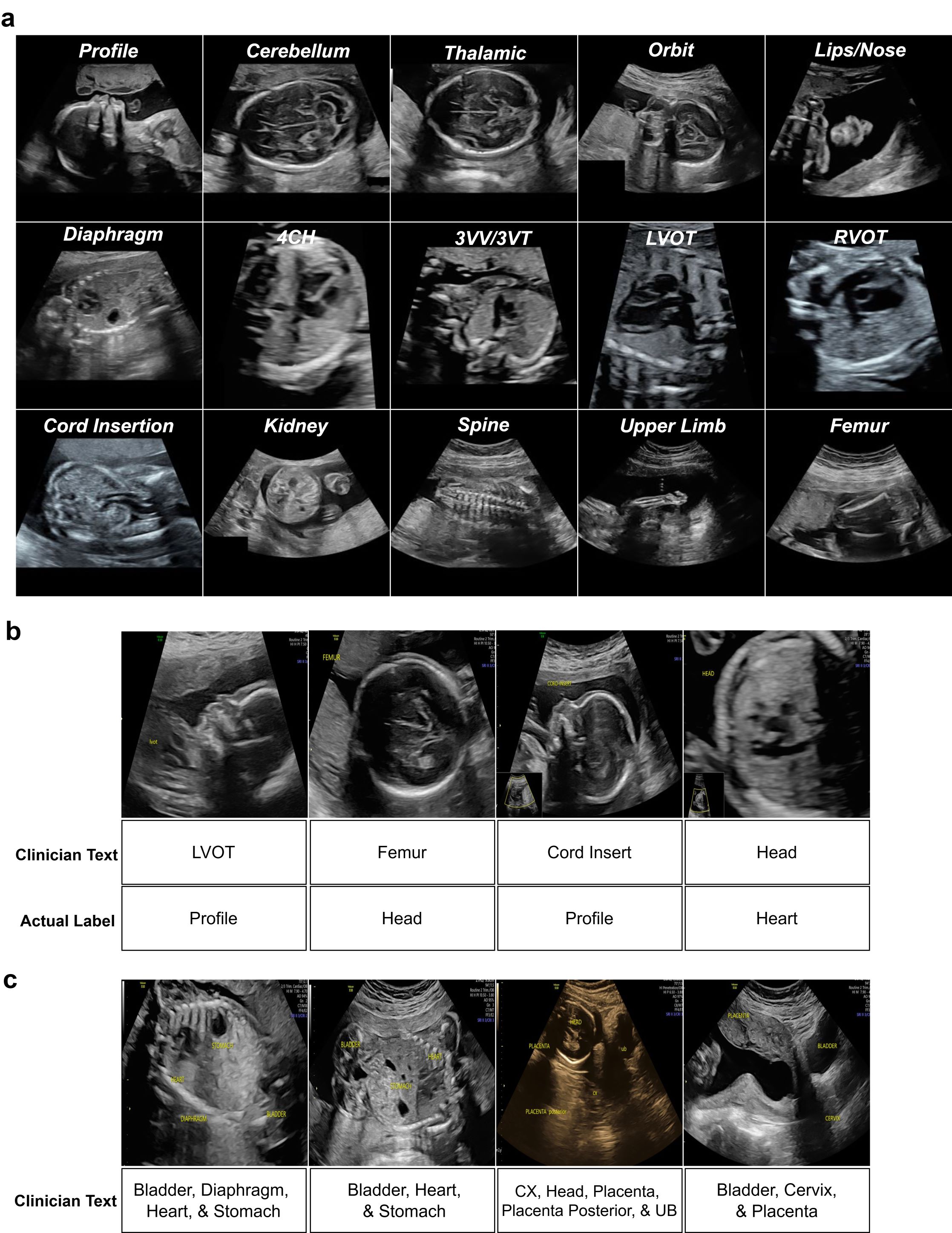}
    \caption{\textbf{Examples of various image views from the private hospital dataset.}
        \textbf{a,} Representative examples of standard views from the fetal ultrasound dataset, showcasing diverse anatomical planes such as 4CH, Femur, Kidney, and Transcerebellum.
        \textbf{b,} Examples of mislabeled samples detected by Confident Learning.
        \textbf{c,} Ultrasound images containing multiple clinician labels. 
    }
    \label{ext_fig_corniche_dataset}
\end{extended_figure}

\newpage
\begin{extended_figure}[ht!]
    \centering
    \includegraphics[width=\linewidth]{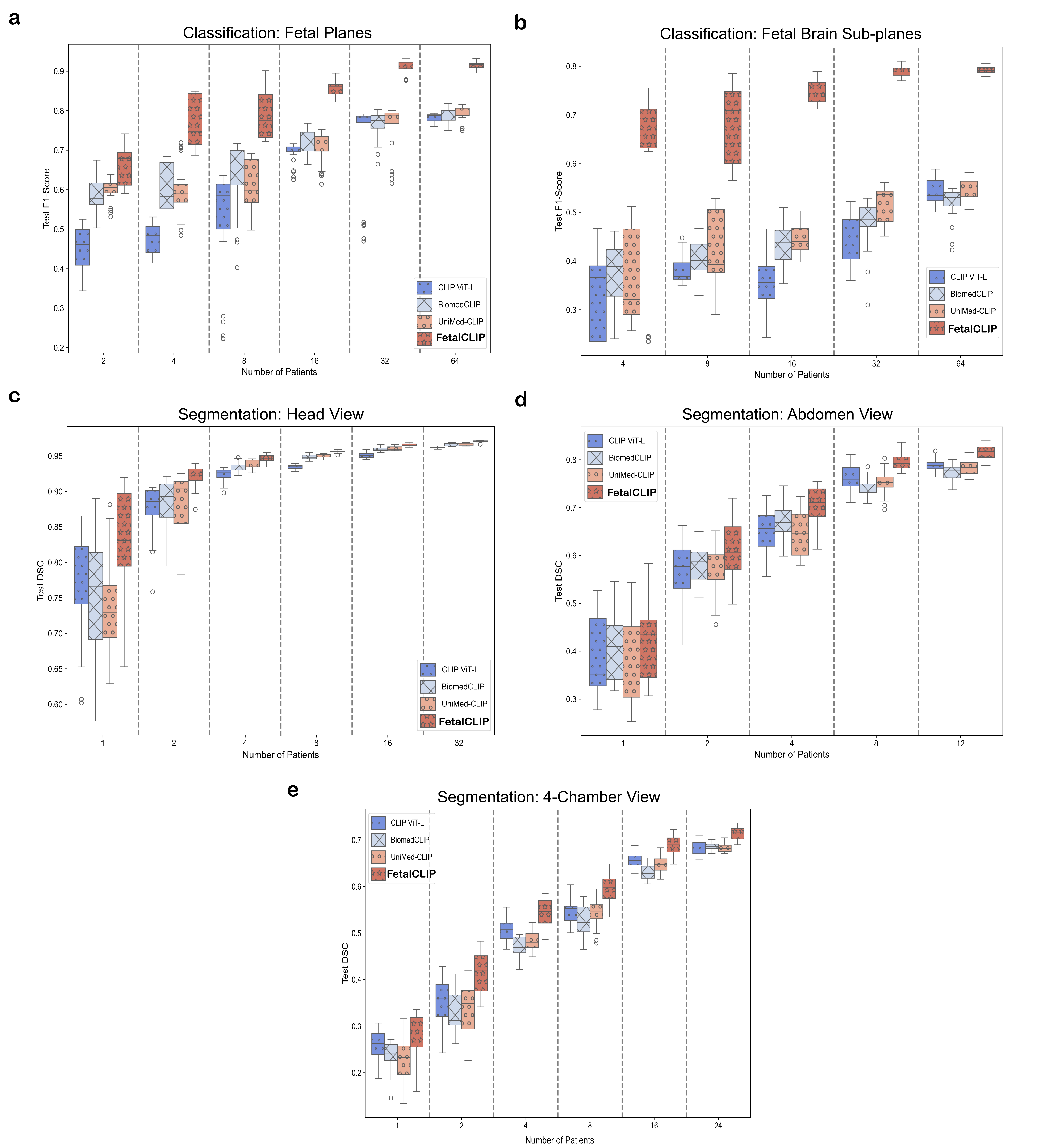}
    \caption{
        \textbf{Data-efficient transfer learning for downstream tasks.} The image encoder was kept frozen during these experiments. Data from $N$ patients were used for training, with an additional $N$ patients for the validation set, collectively defined as the support set, while the final performance was evaluated on an independent test set. Experiments were conducted across five support sets using five different random seeds, yielding a total of 25 results. \textbf{a-b}, Classification performance for six fetal planes and three brain subplanes, respectively. \textbf{c–e}, Segmentation performance on the head, abdomen, and four-chamber views, respectively.
    }
    \label{fig_few_patients}
\end{extended_figure}

\newpage
\begin{extended_figure}[ht!]
    \centering
    \includegraphics[width=\linewidth]{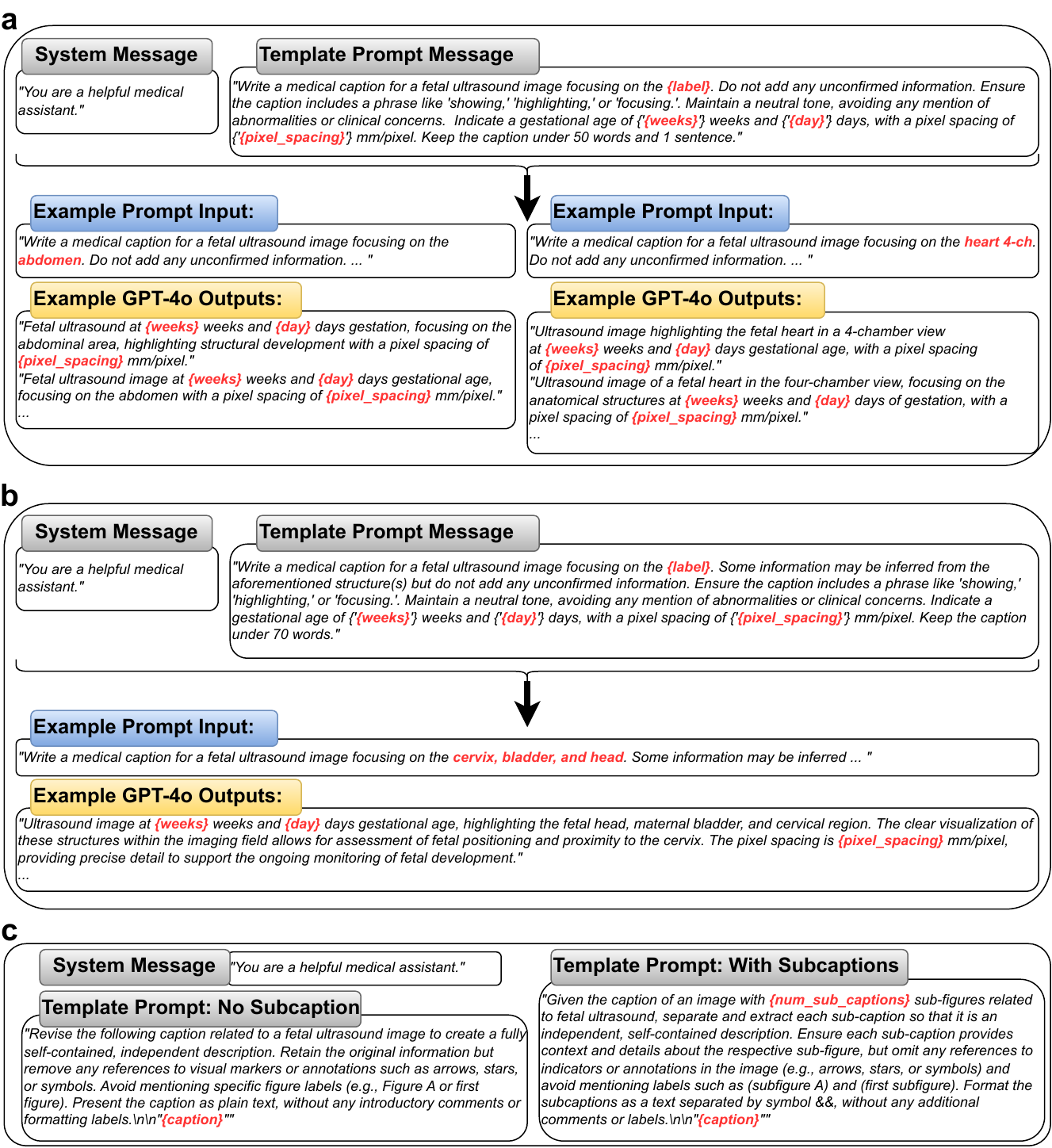}

    \caption{
        \textbf{Prompts used to generate captions for routine clinical scan data and image-caption pairs derived from a textbook.}
        \textbf{a,} Prompts for generating caption templates for the 12 standard views, four cardiac subviews, and brain subplanes.
        \textbf{b,} Prompts for generating caption templates for other diverse labels, including images with multiple views.
        \textbf{c,} Prompts for preprocessing captions to generate subcaptions for datasets derived from image-caption pairs from a textbook.
    }
    \label{ext_fig_prompts_training}
\end{extended_figure}

\newpage
\begin{extended_figure}[ht!]
    \centering
    \includegraphics[width=\linewidth]{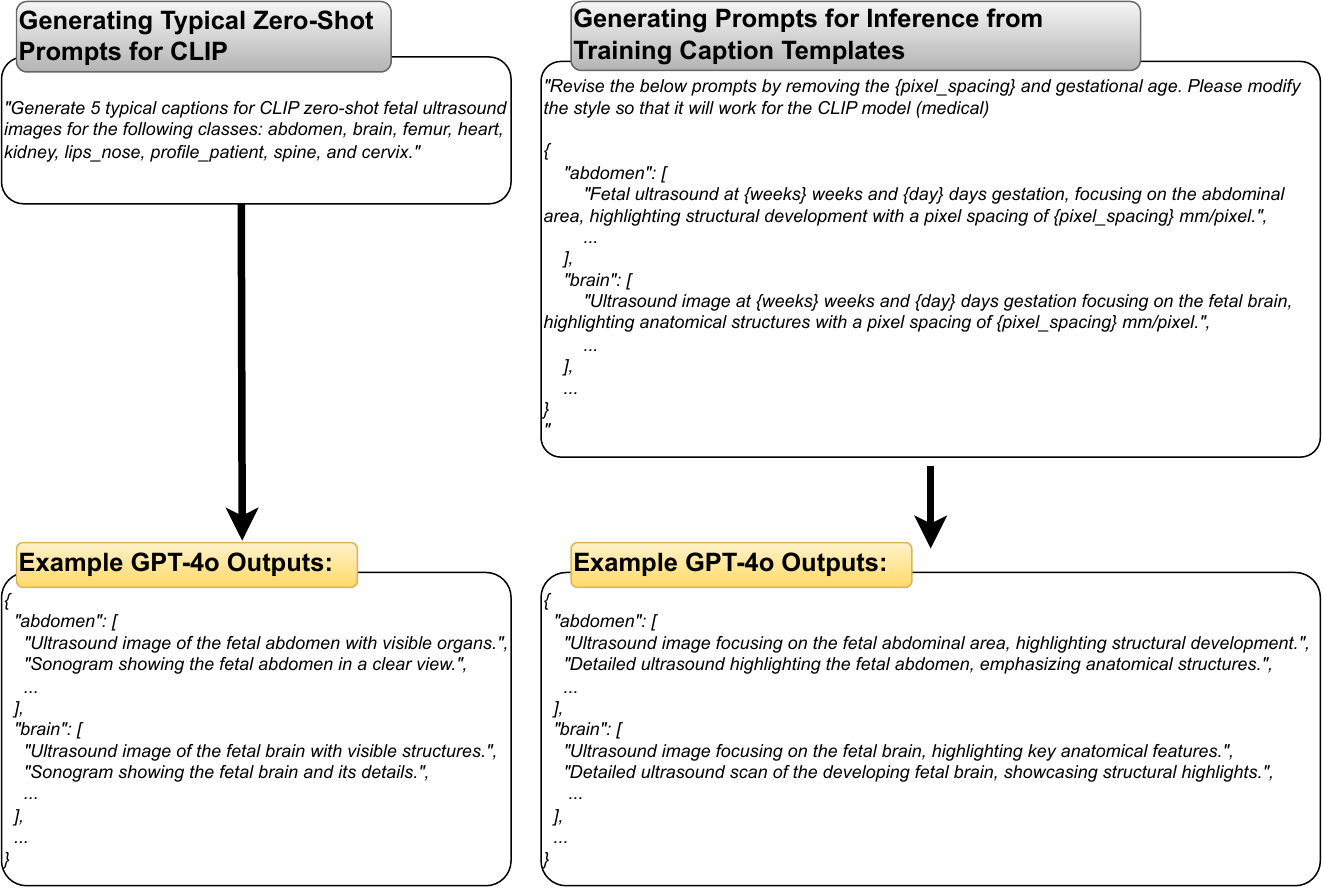}
    \caption{
        \textbf{Prompts for inference.}
        Five text prompts were designed for each target class using GPT-4o. We generated two types of inference prompts: typical prompts for CLIP models and prompts incorporating information from caption templates of routine pregnancy scan data.
    }
    \label{prompts_for_inference}
\end{extended_figure}

\newpage
\begin{extended_table}[ht!]
    \caption{\textbf{Dataset details and training protocols for downstream tasks.} Two public datasets (Planes DB \cite{planes_db}, HC18 \cite{hc18}) and three private datasets were utilized for benchmarking. Two experimental settings were investigated: full data and data-efficient training. Image augmentation was performed offline and implemented using Albumentations \cite{albumentations}.}
    \begin{tabular}{c}
        \includegraphics[width=0.95\linewidth]{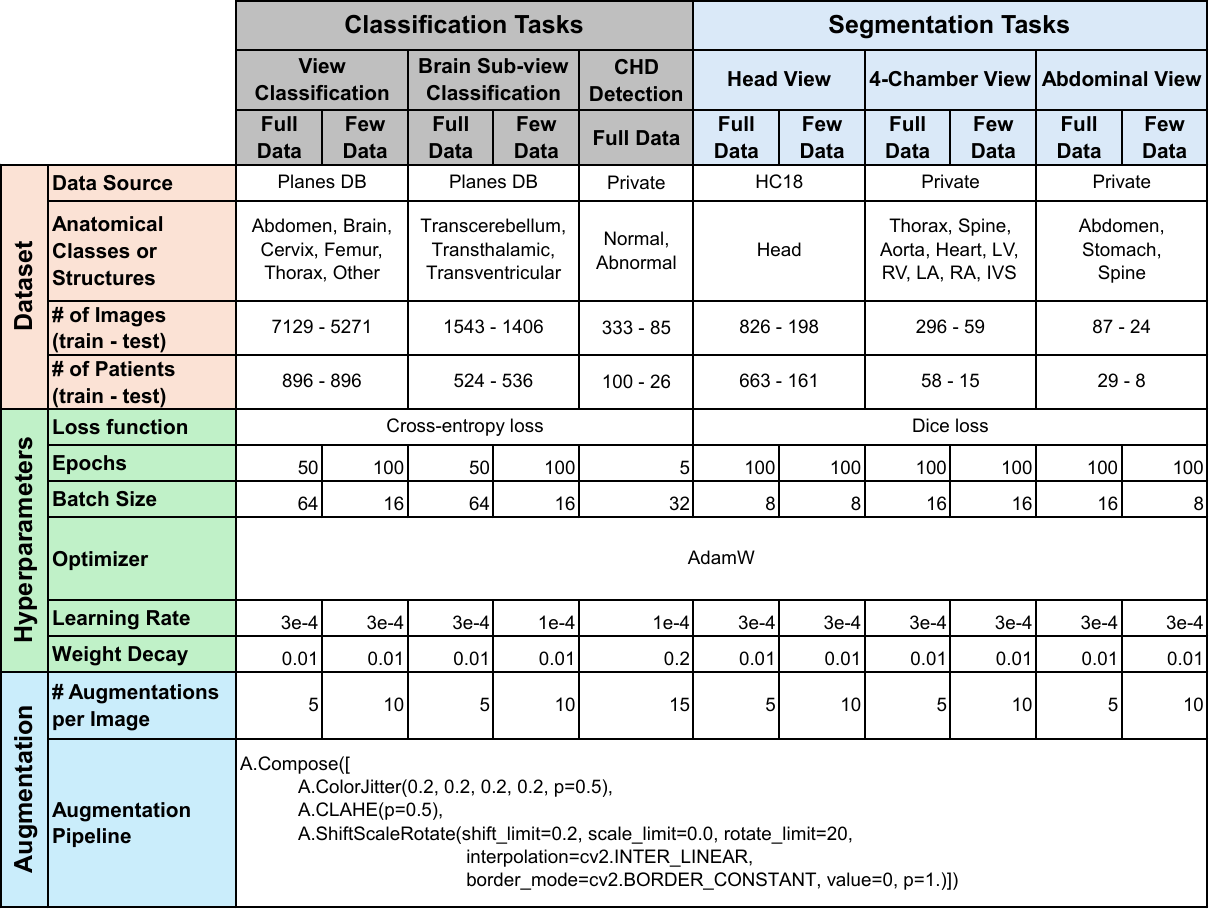}
    \end{tabular}
    \label{tab:ext_tab_downstream_setting}
\end{extended_table}

\end{document}